\documentstyle[11pt,epsfig]{article}
\newcommand{\bee}   {\begin{equation}}
\newcommand{\ene}   {\end{equation}}
\newcommand{\beqa}  {\begin{eqnarray}}
\newcommand{\enqa}  {\end{eqnarray}}
\newcommand{\bea}   {\begin{array}}
\newcommand{\ena}   {\end{array}}
\newcommand{\dy} {\displaystyle}

\newenvironment{goodlist}[1]
{\begin{list}{}{\settowidth{\labelwidth}{#1}
  \setlength{\leftmargin}{\labelwidth}
  \addtolength{\leftmargin}{\labelsep}
  \setlength{\parsep}{1ex plus0.7ex minus0.7ex}
  \setlength{\itemsep}{0.8ex}
  }}{\end{list}}

\def\ga{\gamma}
\def\als{{\alpha_s}}
\def\upa{\uparrow}
\def\dna{\downarrow}
\def\xdqi{x\Delta q^i}

\def\dqit{\Delta\tilde q^i}
\def\dgt{\Delta\tilde{g}}
\def\alp{\frac{\als(Q^2)}{\pi}}
\def\ald{3.58}
\def\alt{20.22}

\font\ten=cmbx10 at 12pt

\begin{document}

\thispagestyle{empty}
\baselineskip=18pt

\begin{titlepage}
\null
\vspace{-1.5truecm}
\begin{center}
{\ten Centre de Physique Th\'eorique\footnote{Unit\'e Propre de
Recherche 7061} - CNRS - Luminy, Case 907}
{\ten F-13288 Marseille Cedex 9 - France }
\vspace{.9truecm}

{\Large\bf QUARKS AND GLUONS IN NUCLEON 
\\ POLARIZED STRUCTURE FUNCTIONS\footnote{Talk given by J. Soffer at the
Conference on the Fundamental Structure of Matter, Ouranoupoulis (Greece)
May 28-31 (1997).}}

\vspace{0.1truecm}

\noindent
~\\
{\bf C. BOURRELY$^a$, F. BUCCELLA$^{b,c}$, O. PISANTI$^{b,c}$, 
P. SANTORELLI$^{b,c}$}\\
and \\
{\bf J. SOFFER$^a$}
~\\
~\\
\begin{it}
$~^a$Centre de Physique Th\'eorique--CNRS
Luminy, Case 907\\
F-13288 Marseille Cedex 9, France\\
$~^b$Dipartimento di Scienze Fisiche, Universit\`a ``Federico II'',\\
Pad. 19 Mostra d'Oltremare, 00195 Napoli, Italy \\
$~^c$INFN, Sezione di Napoli,\\
Pad. 20 Mostra d'Oltremare, 00195 Napoli, Italy
\end{it}

\vspace{0.5 cm}
{\bf Abstract}
\end{center}
We study the available data in polarized $e-p$ deep inelastic scattering to
test two different solutions to the so called {\it spin crisis}: one of
them based on the axial gluon anomaly and consistent with the Bjorken sum
rule and another one, where the defects in the spin sum rules and in the
Gottfried sum rule are related. In this case a defect is also expected for
the Bjorken sum rule. Experimental data, especially the very recent SLAC
E154, favour the first solution and demand a gluon polarization $\Delta G =
2.25\,\pm\,1.39$.

\bigskip

\noindent Number of figures : 5 \\
\noindent keywords : polarized parton distributions, sum rules.\\
\noindent March 1996 (Revised December 1996) \\
\noindent DSF$-$T$-$96/17 \\
\noindent CPT$-$96/PE.3327 \\
\noindent anonymous ftp or gopher: cpt.univ-mrs.fr
\end{titlepage}

\section{Introduction}

The earlier EMC CERN experiment \cite{emc} and the importance of testing
the Bjorken sum rule \cite{bjork} have stimulated a considerable
experimental activity in measuring $g_1^p$, $g_1^n$ ($g_1^{He^3}$) and
$g_1^d$. The result found from EMC for the first moment,
\bee
\Gamma_1^p = \int_0^1 g_1^p(x)~ dx = 0.126\pm 0.010\pm 0.015,
\label{e:gapemc}
\ene
has been confirmed by the SMC CERN experiment \cite{smcp} (at
$<\!\!Q^2\!\!> = 10\,GeV^2$, which is almost the same as for EMC) and SLAC
\cite{slacp} (at $<\!\!Q^2\!\!> = 3\,GeV^2$), giving respectively
\bee
\bea{cccc}
\Gamma_1^p &=& 0.136\pm 0.011\pm 0.011 & (SMC), \\[.3cm]
\Gamma_1^p &=& 0.127\pm 0.004\pm 0.010 & (E143).
\ena
\label{e:gapsmcslac}
\ene
These experiments have also measured the deuteron structure function
\cite{smcd,slacd} and from $\Gamma_1^d$ by substracting $\Gamma_1^p$, one
gets
\bee
\bea{cccc}
\Gamma_1^n &=& -0.063\pm 0.024\pm 0.013  & (SMC), \\[.3cm]
\Gamma_1^n &=& -0.037\pm 0.008\pm 0.011  & (E143).
\ena
\ene
At SLAC with polarized $H\!e^3$ targets they also obtained \cite{e142}
\bee
\Gamma_1^n = -0.031\pm 0.006\pm 0.009 \quad\quad\quad\quad\quad\quad\quad
(E142, \,Q^2= 2 GeV^2), 
\ene
and the preliminary result \cite{e154}
\bee
\int_{0.014}^{0.7} g_1^n (x) dx = -0.037 \pm 0.004 \pm 0.010 \quad\quad
(E154, \,Q^2= 5 GeV^2).
\ene
The preliminary result from Hermes \cite{hermes} ($\Gamma_1^n = -0.032\,
\pm\, 0.013\, \pm\, 0.017$ at $<\!\!Q^2\!\!> = 3\,GeV^2$) is consistent
with SLAC data.

The main issue of this experimental work is to test the validity of the
Bjorken sum rule, which up to $O (\als^3)$, for $n_f = 3$, is given by
\cite{correc}
\bee
\Gamma_1^p - \Gamma_1^n = \frac{1}{6} \frac{G_A}{G_V} \left[ 1 - \alp -
\ald \left( \alp \right)^2 - \alt \left( \alp \right)^3 \right]. 
\ene
Indeed, an interpretation of the defect in the Ellis and Jaffe sum rule
\cite{elljaf} for $\Gamma_1^p$ implied by Eqs.~(\ref{e:gapemc}) and
(\ref{e:gapsmcslac}) has been given in terms of a negative contribution
coming from a large positive polarization of the gluons $\Delta G$
\cite{anom}, which is the same for proton and neutron, such that it does
not affect the Bjorken sum rule.

An analysis of the existing data, excluding the most recent and very
precise SLAC E154 data, has been performed in a framework consistent with
the Bjorken sum rule and including next to leading order (NLO) effects in
the evolution equations to relate data at different $Q^2$, and it provides
a fair description of the experimental results \cite{nlofit}.

All the existing data do not exhibit a clear evidence of $Q^2$ evolution,
i.e. within error bars they are compatible with scaling behaviour in the
entire $Q^2$ range accessible by all experiments. Although a complete NLO
analysis of the $g_1$ data is the correct procedure without a doubt, we are
aiming to demonstrate that the most accurate data provided by the SLAC
experiment is really telling us something important about the gluon
contribution, even at a level of a less sophisticated leading order (LO)
analysis. In spite of the fact that the SMC CERN experiment can reach
higher $Q^2$ and smaller $x$, the actual precision achieved cannot provide
a reliable test for the NLO theoretical analysis.

Here we want to compare the current interpretation of the defect in the
Ellis and Jaffe sum rule in terms of a large flavour singlet contribution
to the nucleon polarization coming from the gluons, with another one
\cite{pauli}, where one relates this defect to the one in the Gottfried sum
rule \cite{gottfr} and to the role that Pauli principle seems to play
\cite{fielfeyn}. This is done by relating the first moments and the shapes
of the parton distributions, as first proposed in Ref.~\cite{pauli}.
Indeed, if one assumes the validity of the Adler sum rule \cite{adler},
\bee
u - d = \left[ (u -\bar u) - (d - \bar d) \right] + \bar u - \bar d =
1 + \bar u - \bar d,
\ene
the defect in the Gottfried sum rule implies
\bee
u - d < 1.
\label{e:mored}
\ene
If one thinks that the Pauli principle is responsible for the inequality
(\ref{e:mored}), it is reasonable to assume that it is $u^\upa$, the most
abundant valence parton, which receives less contribution from the sea, so
that we have \cite{pauli}
\bee
\Delta u = u^\upa - u^\dna \simeq  \Delta u_{val} + \bar u - \bar d,
\ene
producing a defect in the Ellis and Jaffe sum rule for the proton
\bee
\Delta \Gamma_1^p = \frac{2}{9}\, (\bar u - \bar d) = \frac{2}{9}\, (-0.15
\pm 0.04) \simeq - 0.033 \pm 0.009,
\label{e:dg}
\ene
in fair agreement with the experiments.

An empirical test for the two interpretations might be given using the
experimental information on the $x$-dependence of the polarized structure
functions. The higher precision of SLAC data (especially the ones of E154
for $g_1^n$) and the agreement between E142 \mbox{($<\!\!Q^2\!\!> =
2\,GeV^2$)} and E154 ($<\!\!Q^2\!\!> = 5\,GeV^2$) data, suggest to describe
them together with the E143 measurements ($<\!\!Q^2\!\!> = 3\,GeV^2$) on
proton and deuteron in terms of the same parton distributions, and we
consider for these distributions two options corresponding to the two
different solutions to the {\it spin crisis}.

We shall neglect higher-twist terms, supported by more recent theoretical
evaluations of the contribution of these terms \cite{hightwi} which lead to
results smaller in modulus and sometimes opposite in sign than the previous
one, consistent with an experimental determination of these terms by the
SLAC group \cite{hightwisp}.

The paper is organized as follows. In the forthcoming section we shall
describe the SLAC data, with proton and deuteron targets at $<\!\!Q^2\!\!>
= 3\,GeV^2$ and $He^3$ target at $<\!\!Q^2\!\!>= 2\,GeV^2$ (E142) and at
$<\!\!Q^2\!\!>= 5\,GeV^2$ (E154), with the two different options. Then we
shall present the method we used to solve the Altarelli-Parisi evolution
equations and to find the parton distributions at $Q^2 = 10\,GeV^2$ which
we shall compare with CERN data. Finally, we shall give our conclusions.

\section{Description of SLAC data}

We describe the proton and neutron polarized structure functions at $Q_0^2
= 3\,GeV^2$, in terms of the valence quark and gluon polarized
distributions only, using a simplified version of the functional forms used
in Ref.\cite{gerst} (in our case we take $\ga_q (q = u,d,G) = 0$), namely
\bee
\bea{lcl}
x \Delta u_v (x, Q_0^2) &=& \eta_u A_u x^{a_u} (1-x)^{b_u},
\vspace{.2truecm} \\ 
x \Delta d_v (x, Q_0^2) &=& \eta_d A_d x^{a_d} (1-x)^{b_d},
\vspace{.2truecm} \\
x \Delta G (x, Q_0^2)   &=& \eta_G A_G x^{a_G} (1-x)^{b_G},
\ena
\label{e:distr}
\ene
where $\eta_q$ ($q~=~u,d,G$) are the first moments of the distributions and
$A_q = A_q(a_q,b_q$),
\bee
A_q^{-1} =\int_0^1 dx x^{a_q - 1} (1-x)^{b_q} = \frac{\Gamma(a_q)
\Gamma(b_q+1)}{\Gamma(a_q+b_q+1)},
\ene
in such a way that
\bee
\int_0^1 dx A_q x^{a_q-1} (1-x)^{b_q} = 1.
\ene

As pointed out by several authors \cite{scheme}, to avoid the inclusion of
soft contributions into the coefficient functions one has to choose a
factorization scheme in which the gluon polarization contributes to the
first moments of $g_1^p$ and $g_1^n$ (for $n_f = 3$):
\bee
\Gamma_1^{p(n)}(Q^2) = \frac{2}{9} \left( \frac{1}{18} \right) \eta_u(Q^2)
+ \frac{1}{18} \left( \frac{2}{9} \right) \eta_d(Q^2) -
\frac{\als(Q^2)}{6\, \pi} \eta_G(Q^2).
\ene
The gluonic term appears to be a higher order correction but is not,
because $\eta_G(Q^2)$ rises logarithmically with $Q^2$ and, if the gluons
had a positive polarization, it could, in principle, be large enough to
explain the defect in the Ellis and Jaffe sum rule.

We take
\bee
\bea{rcl}
g_1^{p(n)} (x,Q^2) &=& \dy \frac{2}{9} \left( \frac{1}{18} \right) \Delta
u_v (x,Q^2) + \frac{1}{18} \left( \frac{2}{9} \right) \Delta d_v (x,Q^2) -
\frac{\als (Q^2)}{6\,\pi} \left( \Delta\sigma \otimes \Delta G \right)
(x,Q^2), \\[.5cm]
g_1^d (x,Q^2) &=& \dy \frac{1}{2} \left( 1 - \frac{3}{2} \omega_D \right)
(g_1^p(x,Q^2) + g_1^n(x,Q^2)),
\ena
\label{e:g1pnd}
\ene
where $\omega_D = 0.058$ \cite{omD} takes into account the small D-wave
component in the deuteron ground state. In Eq.~(\ref{e:g1pnd}) the QCD
corrections in the quark sector are included in the $\tilde F$ and $\tilde
D$ values entering in the expressions of the first moments of the quark
distributions ($F = 0.46 \pm 0.01$, $D = 0.79 \pm 0.01$ \cite{hyper}):
\bee
\bea{rcl}
\tilde F(Q^2) &=& \dy \frac{1}{5} \left[ 5\,F \left(1 - \alp\right) -
(10.46 F + 2.48 D)\, \left(\alp\right)^2 \right. \\[.4cm]
&& \dy \left. -~ 20.22~ (2\,F + D)\, \left(\alp\right)^3 \right], \\[.4cm]
\tilde D(Q^2) &=& \dy \frac{1}{5} \left[ 5\,D \left(1 - \alp\right) -
(7.44 F + 15.42 D)\, \left(\alp\right)^2 \right. \\[.4cm]
&& \dy \left. -~ 20.22~ (3\,F + 4\,D)\, \left(\alp\right)^3 \right],
\ena
\ene
and the gluon contribution appears as a convolution \cite{conv}, 
\bee
\left( \Delta\sigma \otimes \Delta G \right) (x,Q^2) = \int_x^1
\frac{dz}{z}\, (1-2\,z) \left( ln \frac{1-z}{z} - 1 \right)\, \Delta G
\left(\frac{x}{z}, Q^2\right).
\ene 
We fix ($\als (3\,GeV^2) = 0.35 \pm 0.05$)
\bee
\eta_d(Q_0^2) = \tilde F(Q_0^2) - \tilde D(Q_0^2) = -0.26\pm 0.02,
\label{e:relat}
\ene
and we explore the two options {\bf A} and {\bf B}, the first one with
\bee
\eta_u(Q_0^2) = 2 \tilde F(Q_0^2) = 0.76\pm 0.04,
\ene
and $\eta_G$ free, the second one with $\eta_u$ free and $\eta_G = 0$.
Options {\bf A} and {\bf B} correspond respectively to the interpretation
of the defect in the Ellis and Jaffe sum rule for $\Gamma_1^p$ in terms of
the anomaly, assuming that the Bjorken sum rule is obeyed, and to the case
of a smaller $\Delta u$ resulting from the Pauli principle.

Since we know that $u^\upa$ dominates at high $x$ and that the gluons
dominate in the small $x$ region, we restrict, as in Ref.\cite{anomaly},
the values of the parameters in Eqs.~(\ref{e:distr}), to be consistent with
the information we already have for the parton distributions, by the
following limitations
\bee
b_u > 1, ~~~~~b_d > 3, ~~~~~b_G > 5,
\ene
and we assume
\bee
a_u = a_d.
\ene
Indeed, especially for option {\bf A}, where one describes two functions,
$g^p_1(x)$ and $g^n_1(x)$, in terms of three distributions,
Eqs.~(\ref{e:distr}), one has to make sure to exclude some choices of the
parameters describing well the data, but not consistent with the
information one has from the unpolarized data, that is, {\it e.g.} about
1/2 of the proton momentum (in the $P_z = \infty$ frame) is carried by the
gluons and that the partons $u^\upa$ are dominating the high $x$ region.

The parameters corresponding to the best fit of the SLAC proton and
deuteron data for options {\bf A} and {\bf B} are given in Table
\ref{t:fitaeb}, while in Figs.~1, 2 and 3 one compares the two resulting
curves with SLAC data.

All the data are well described with the two options, except for the ones
by E154, which are better described by the option {\bf A}, the one with
gluon contribution and consistent with the Bjorken sum rule. Option {\bf A}
implies a large value of $\Delta G = 2.25 \pm 1.39$.

\section{Parton evolution equations}

For the polarized parton distributions one has the
Dokshitzer-Gribov-Lipatov-Altarelli-Parisi evolution equations (DGLAP)
\cite{dglap}, which are, in the variable $t \equiv \ln Q^2/\Lambda^2_{QCD}$
and at LO in $\als$ ($\dqit\equiv\xdqi$ and $\dgt\equiv x\Delta G$),
\bee
\bea{rcl}
\dy \frac{d\dqit}{dt} (x,t) &=& \dy \frac{\als (t)}{2\pi} \left[ \int_x^1
dz~ \frac{4}{3} \left[ \frac{2}{(1-z)_+} - 1 - z + \frac{3}{2} \delta(1-z)
\right] \dqit \left( \frac{x}{z},t \right) \right.\vspace{.4truecm} \\
+ && \!\!\!\!\!\!\!\!\!\!\! \dy \left. \int_x^1 dz~ \left( z - \frac{1}{2}
\right) \dgt \left( \frac{x}{z},t \right) \right], \quad\quad\quad\quad
\quad\quad\quad\quad (i = 1, ..., 2\,n_f) \vspace{.4truecm} \\
\dy \frac{d\dgt}{dt} (x,t) &=& \dy \frac{\als (t)}{2\pi} \left[ \int_x^1
dz~ \frac{4}{3} (2 - z) \sum_{i=1}^{2\,n_f} \dqit \left( \frac{x}{z},t
\right) \right. \vspace{.4truecm} \\
+ && \!\!\!\!\!\!\!\!\!\!\! \dy \int_x^1 dz~ 3 \left[ \frac{2}{(1-z)_+} +
2  -4\,z + \left( \frac{11}{6}-\frac{n_f}{9} \right) \delta(1-z) \right]
\dgt \left( \frac{x}{z},t \right). \vspace{.4truecm}\\
\ena
\label{e:APpol}
\ene

We work in the fixed flavour scheme where the number of flavours in the
splitting functions is $n_f = 3$, while for the $Q^2$-evolution of $\als$
one has
\bee
\als (Q^2) = \frac{12 \pi }{(33 - 2 n_f)~ \ln \frac{\dy Q^2}{\dy
\Lambda_{QCD}^{(4)~2}}},
\ene
with $\Lambda_{QCD}^{(4)} = 201\, MeV$ to reproduce, with $n_f= 4$, $\als
(3\,GeV^2) = 0.35 \pm 0.05$.

For the solution of the DGLAP equations we have used a method \cite{kwiez}
that consists in expanding the parton distributions $p^i$ into a truncated
series of Chebyshev polynomials,
\bee
p^i(x,t) \rightarrow p^i(\tau(x),t) = \frac{2}{n} ~\sum_{s=0}^{n-1}
  ~\sum_{l=0}^{n-1} ~v_l ~p^i(x_s,t) ~T_l(\tau_s) ~T_l(\tau(x)),
\label{e:pcheb}
\ene
where $T_l$ are the Chebyshev polynomials and
\beqa
\tau(x) &=& \frac{-2 ~\ln x-y_{max}}{y_{max}}, \nonumber \\
y_{max} &=& 4~ \ln 10, \nonumber \\
v_l &=& \left\{ \bea{cc}
		0.5 & l=0 \\
		1   & l>0
		\ena  \right., \\
x_s &=& exp \left[ - \frac{y_{max}}{2}~ (\tau_s + 1) \right], \nonumber \\
\tau_s &=& cos~ \frac{2~s + 1}{2~n}\pi. \nonumber
\enqa

Substituting Eq.~(\ref{e:pcheb}) in Eqs.~(\ref{e:APpol}) gives rise to a
system of coupled differential equations
\bee
\frac{d p^i_k}{dt} (t) = \frac{\als (t)}{2 \pi}~ \sum_{j=1}^7~
\sum_{s=0}^{n-1}~ A^{ij}_{ks}~ p^j_s (t),~~~~~~ (i = 1,...,7;~~ k = 0, ...,
n-1),
\label{e:eqdf}
\ene
in which $p^i_k (t)\equiv p^i(x_k,t)$ are the values of the polarized
distributions $\dqit$ and $\dgt$ in the $n$ points $x_k$ corresponding to
the nodes $\tau_k$ of the Chebyshev polynomials.

With the initial conditions given by the results of the fits {\bf A} and
{\bf B} to the SLAC data we get the evolved distributions at $Q^2 =
10\,GeV^2$.

\section{Comparison of the evolved distributions with experiments}

The predictions for the evolved distributions at $Q^2 = 10\,GeV^2$ are
compared with CERN measurements \cite{smcp,smcd} at $Q^2 = 10\,GeV^2$ in
Figs.~4 and 5. There is a better agreement for option {\bf B} for the
proton (total $\chi^2$ of 10.6 for 12 points to be compared with $\chi^2 =
16.2$ for option {\bf A}) while for the deuteron the option {\bf A} has a
slightly lower $\chi^2$ (total $\chi^2$ of 14.2 for 12 points to be
compared with $\chi^2 = 16.8$ for option {\bf B}). Note that for option
{\bf A} one has six free parameters, but only four for option {\bf B}.

It is interesting to remark that with both options one fails to reproduce
the rise at low $x$ of $x g_1^p (x)$, which turns negative below $x =
10^{-2}$ for option {\bf A}, while $x g_1^d (x)$ in the same region is in
agreement with the trend of the data (see Fig.~5). This is due to the
isoscalar nature of the anomaly contribution, which is expected to be the
same for proton, neutron and deuteron (neglecting the small correction
coming from the D-wave component in its ground state).

\section{Conclusions}

We have studied the precise SLAC data on polarized structure functions with
the main purpose of testing the Bjorken sum rule and the necessity of a
gluon contribution to the polarized structure functions. As a result we
find a better description of the data, especially of the very precise ones
by the SLAC E154, with a gluon contribution and imposing the Bjorken sum
rule than for the option without gluons and with the first moment of
$\Delta u$ free. The best fit in the first case is obtained with a rather
large value of $\Delta G = 2.25 \pm 1.39$.

The $\Delta G$ distribution appears to be more singular than the $\Delta
q$'s, leading to the conclusion that $g_1^p (x)$ should also become
negative at small $x$. These results are in agreement with what was found
in Ref.~\cite{softer}. Although we have found here $a_G = 0.44$ and $a_G =
0.13$ in \cite{softer}, these two different powers lead to comparable
values for $\int_{0.01}^{0.1} \Delta G(x)~ dx$, which is 1.5 in
Ref.~\cite{softer} and 1.0 here. Note that in Ref.~\cite{softer}, since the
$(1-x)$ terms were omitted, one expects to get, for gluons and for valence
quarks, a slightly more singular behaviour when $x\rightarrow 0$ than in
the present analysis.

Concerning the test of the Bjorken sum rule, the fact that one gets a
better fit by allowing the presence of a contribution coming from the
gluons speaks in favour of its validity. Assuming the validity of the
Bjorken sum rule, this contribution was advocated to explain the defect of
the Ellis and Jaffe sum rule for the proton. However, the fact that,
according to our description, at low $x$ the isovector contribution is
expected to be overwhelmed by the isoscalar one, which is more singular,
suggests that a precise test of the Bjorken sum rule rather requires very
accurate measurements in the $x$ region where the two contributions are
comparable. Indeed, in the very small $x$ region, where $g_1^p (x)- g_1^n
(x)$ should have a small value being the difference between two almost
equal negative large quantities, normalizations uncertainties can produce
substantial errors. SLAC E154 has measured $g_1^n (x)$ with an outstanding
precision and we look forward to have, in the same $x$ range, a comparable
precision for $g_1^p$ from SLAC E155 \cite{e155}.

\bigskip
\noindent {\bf Acknowledgements}

\noindent This work has been partially supported by the EC contract
CHRX-CT94-0579. We are indebted to E. Hughes for stimulating discussions
and valuable comments.


\newpage

\begin{table}[h]
\begin{center}
TABLE \ref{t:fitaeb}\\
\vspace{.6truecm}
\begin{small}
\begin{tabular}{|c|c|c|} \hline\hline
\rule[-0.4truecm]{0mm}{1truecm}
			& {\bf A}		& {\bf B}		\\
\hline\hline
\rule[-0.4truecm]{0mm}{1truecm}
$a_u=a_d$		& $0.79\pm 0.05^{(*)}$& $1.03\pm 0.07^{(*)}$	\\
\hline
\rule[-0.4truecm]{0mm}{1truecm}
$b_u$			& $1.51\pm 0.17^{(*)}$	& $1.88\pm 0.23^{(*)}$	\\
\hline
\rule[-0.4truecm]{0mm}{1truecm}
$b_d$			& $3.00\pm 0.13^{(*)}$	& $4.52\pm 0.52^{(*)}$	\\
\hline
\rule[-0.4truecm]{0mm}{1truecm}
$a_G$			& $0.44\pm 0.18^{(*)}$	& -			\\
\hline
\rule[-0.4truecm]{0mm}{1truecm}
$b_G$			& $5.00\pm 1.05^{(*)}$	& -			\\
\hline
\rule[-0.4truecm]{0mm}{1truecm}
$\eta_u$		& $0.76\pm 0.04$	& $0.63\pm 0.03^{(*)}$	\\
\hline
\rule[-0.4truecm]{0mm}{1truecm}
$\eta_d$		& $-0.26\pm 0.02$	& $-0.26\pm 0.02$	\\
\hline
\rule[-0.4truecm]{0mm}{1truecm}
$\eta_G$		& $2.25\pm 1.39^{(*)}$	& -			\\
\hline
\rule[-0.4truecm]{0mm}{1truecm}
$\chi_{NDF}^2$		& 0.84			& 1.09			\\
\hline
\end{tabular}
\end{small}
\end{center}
\caption{The results of the options {\bf A} and {\bf B} for the values of
the parameters of the fits at $Q^2 = 3\,GeV^2$. The free parameters are
marked with an asterisk.}
\label{t:fitaeb}
\end{table}

\newpage
\section*{Figure Captions}

\begin{goodlist}{Fig.6}

\item[Fig. 1]
The best fit for the options {\bf A} (solid line) and {\bf B} (dashed line)
(see text) are compared with the SLAC data on proton for $xg_1^p(x)$ at
$<\!\!Q^2\!\!> = 3\,GeV^2$ from Ref.~\cite{slacp}.

\item[Fig. 2]
Same as Fig. 1 for the deuteron SLAC data for $xg_1^d(x)$ at $<\!\!Q^2\!\!>
= 3\,GeV^2$ from Ref.~\cite{slacd}.

\item[Fig. 3]
Same as Fig. 1 for the neutron SLAC data for $xg_1^n(x)$ at $<\!\!Q^2\!\!>
= 2\,GeV^2$ from Ref.~\cite{e142} (triangle) and at $5\,GeV^2$ from
Ref.~~\cite{e154} (boxes).

\item[Fig. 4]
The data on proton for $xg_1^p(x)$ from SMC at $<\!\!Q^2\!\!> = 10\,GeV^2$
from Ref.~\cite{smcp} are compared with the results of the options {\bf A}
(solid line) and {\bf B} (dashed line), evolved to $Q^2 = 10\,GeV^2$.

\item[Fig. 5]
Same as Fig. 4 for the deuteron SMC data for $xg_1^d(x)$ from
Ref.~\cite{smcd}.

\end{goodlist}

\newpage
\pagestyle{empty}

\begin{figure}[p]
\centerline{\epsfig{file=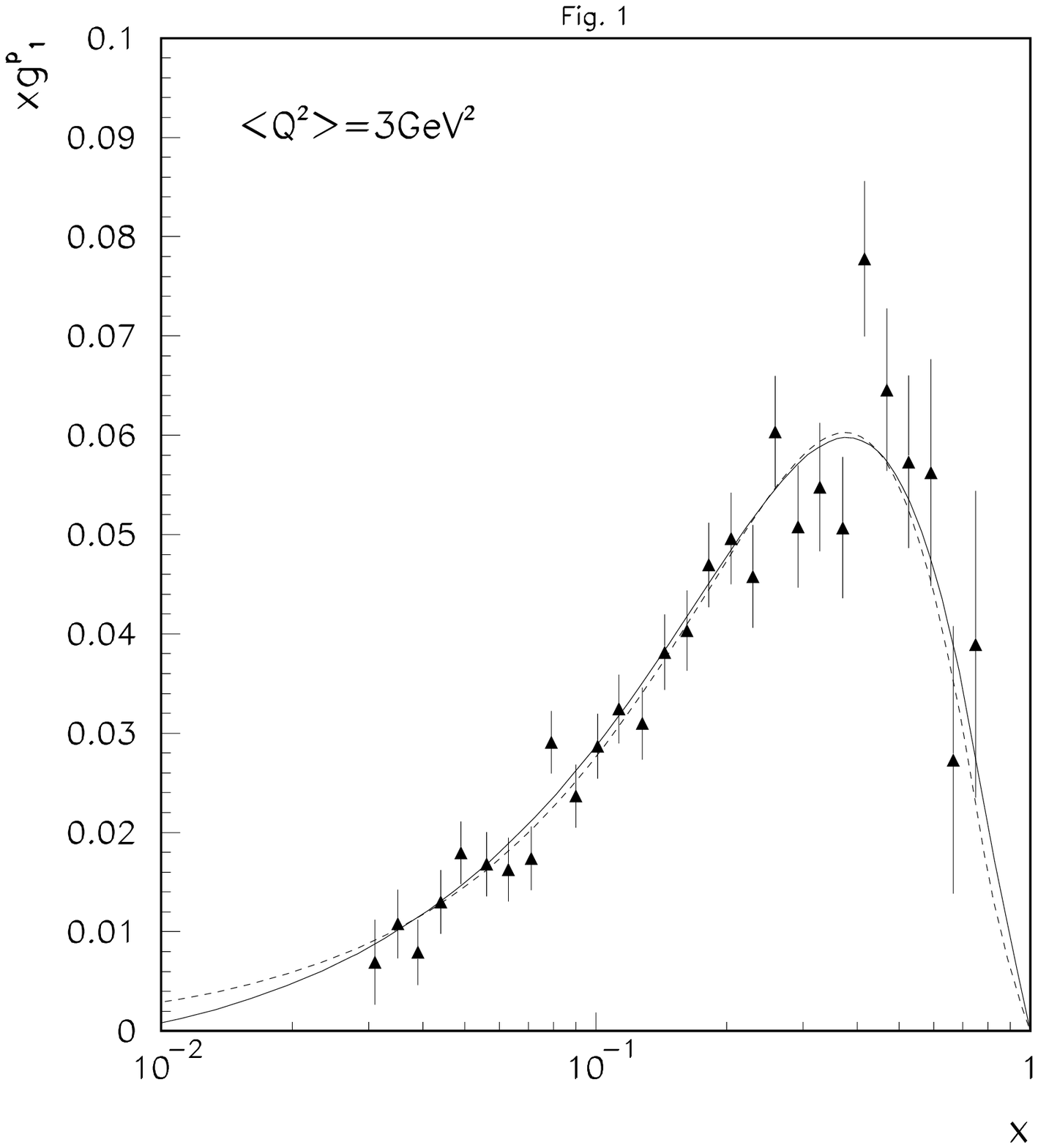,height=15cm}}
\label{fig:fig1}
\end{figure}

\newpage
\begin{figure}[p]
\centerline{\epsfig{file=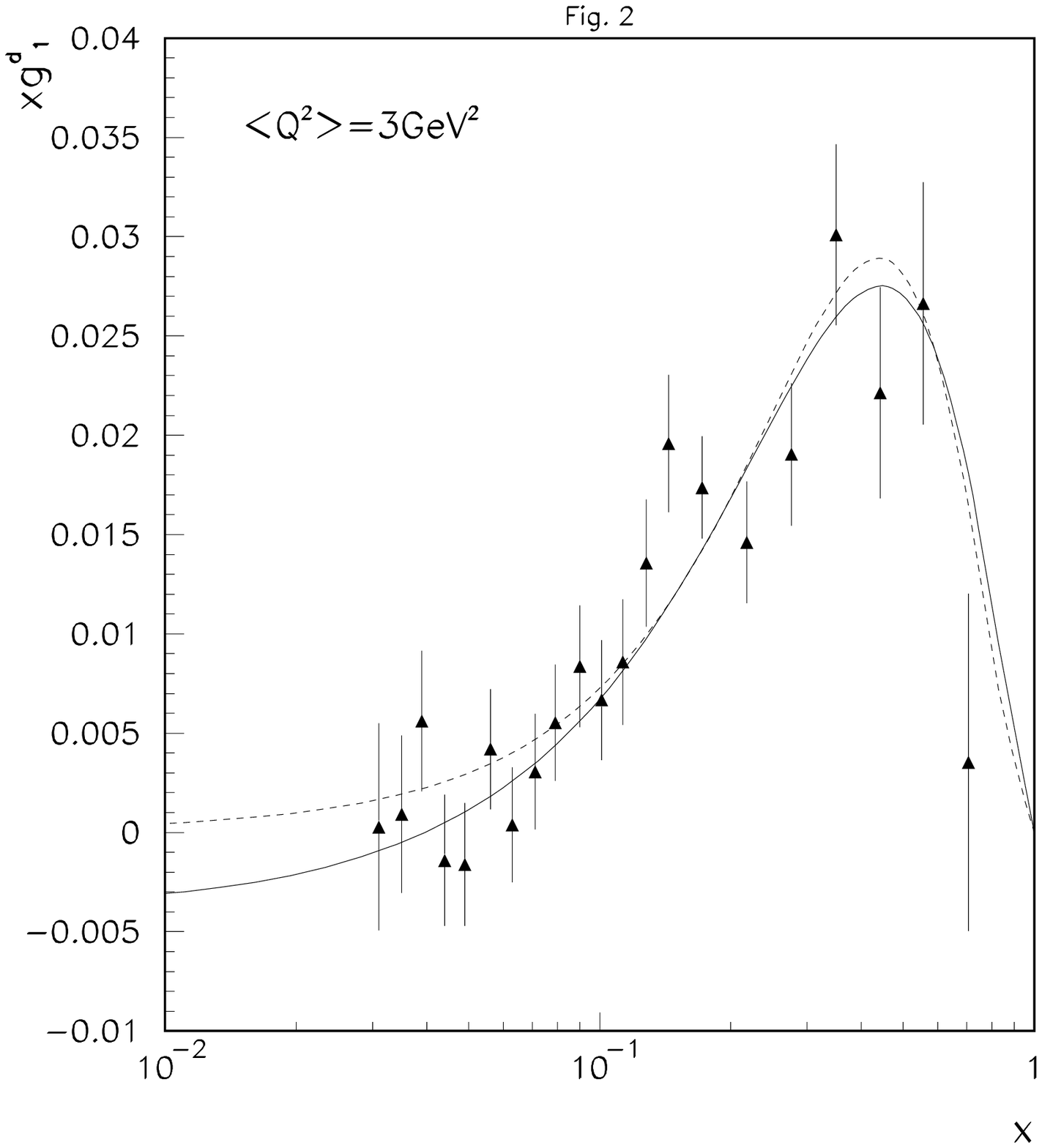,height=15cm}}
\label{fig:fig2}
\end{figure}

\newpage
\begin{figure}[p]
\centerline{\epsfig{file=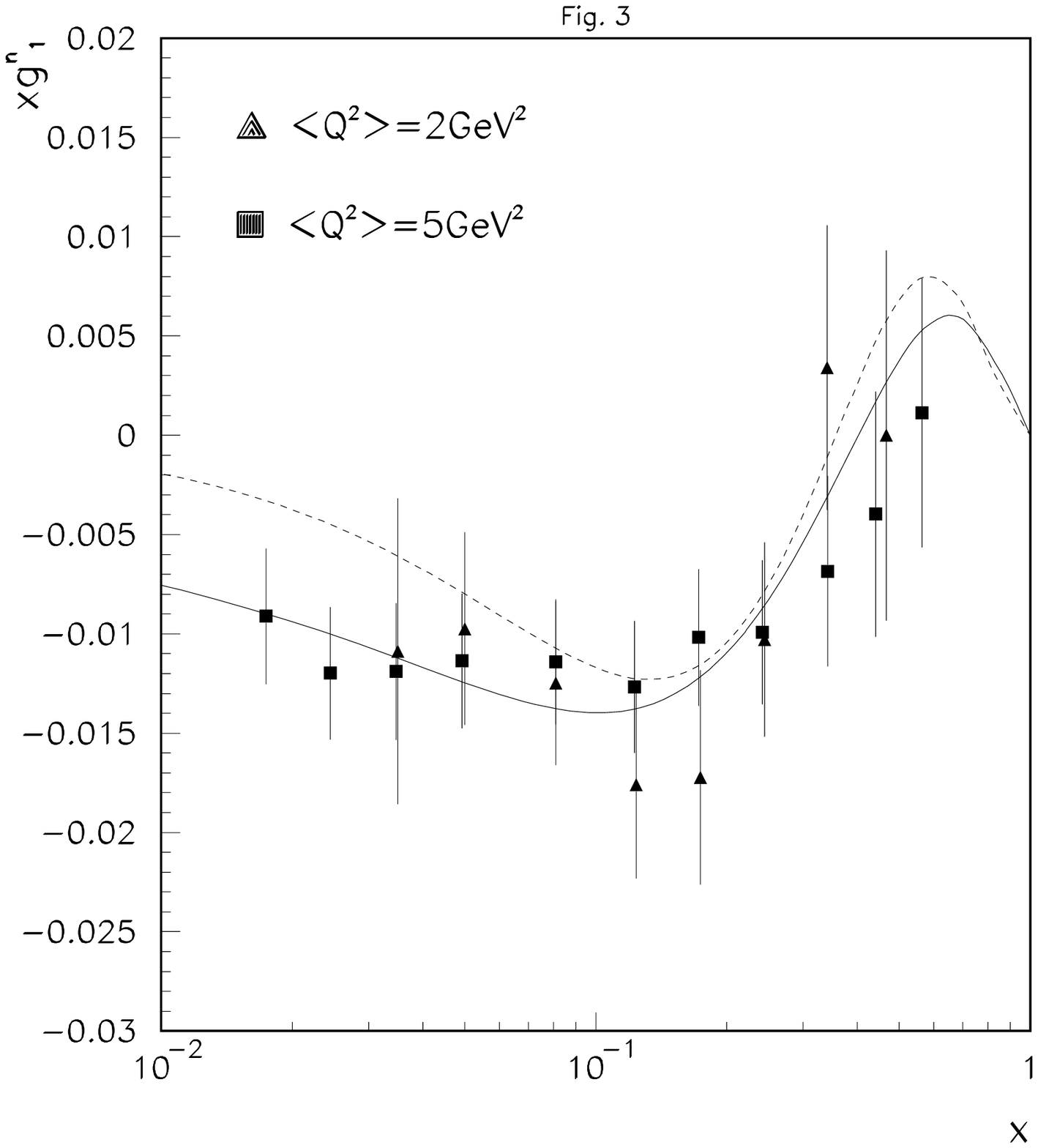,height=15cm}}
\label{fig:fig3}
\end{figure}

\newpage
\begin{figure}[p]
\centerline{\epsfig{file=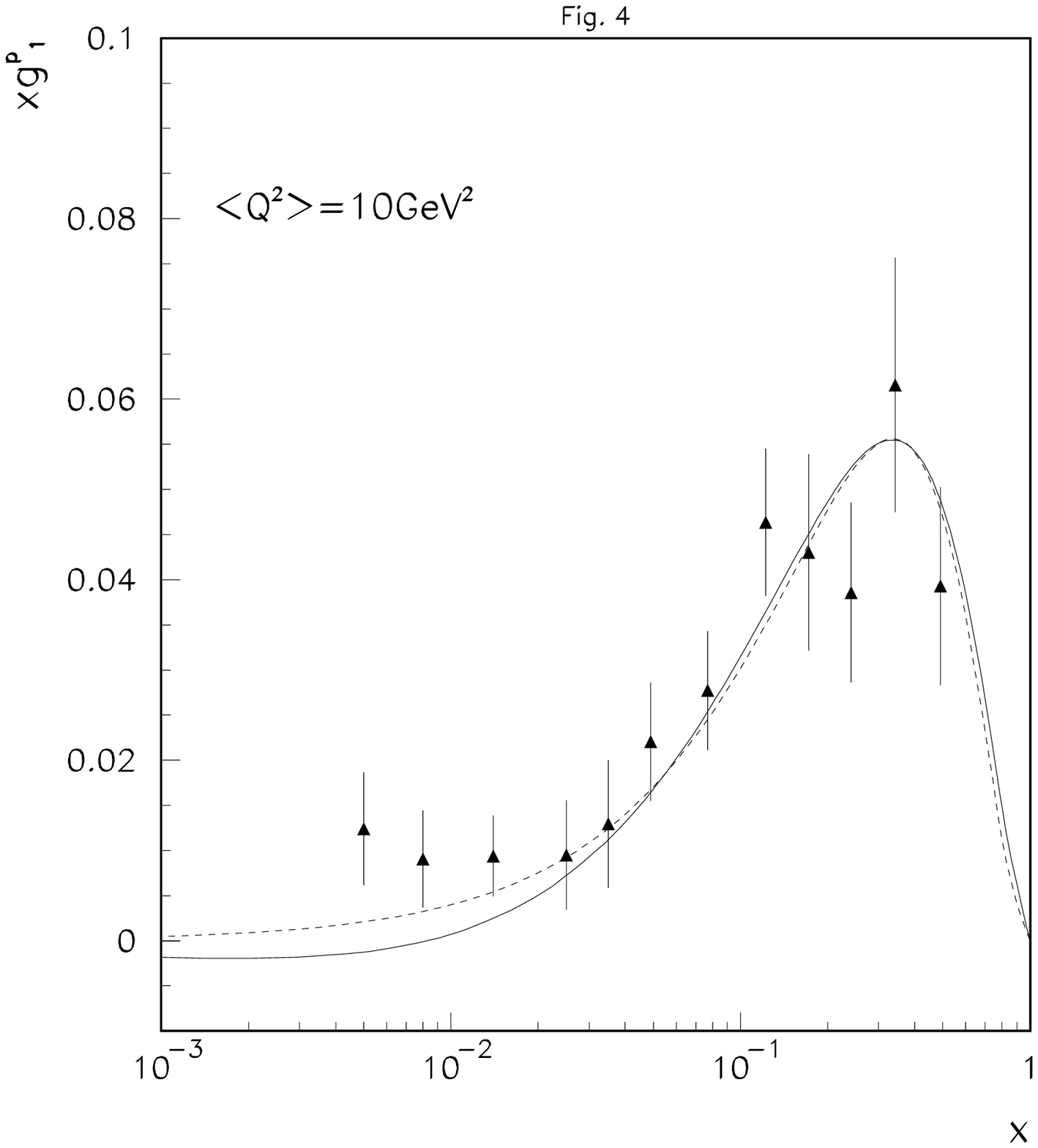,height=15cm}}
\label{fig:fig4}
\end{figure}

\newpage
\begin{figure}[p]
\centerline{\epsfig{file=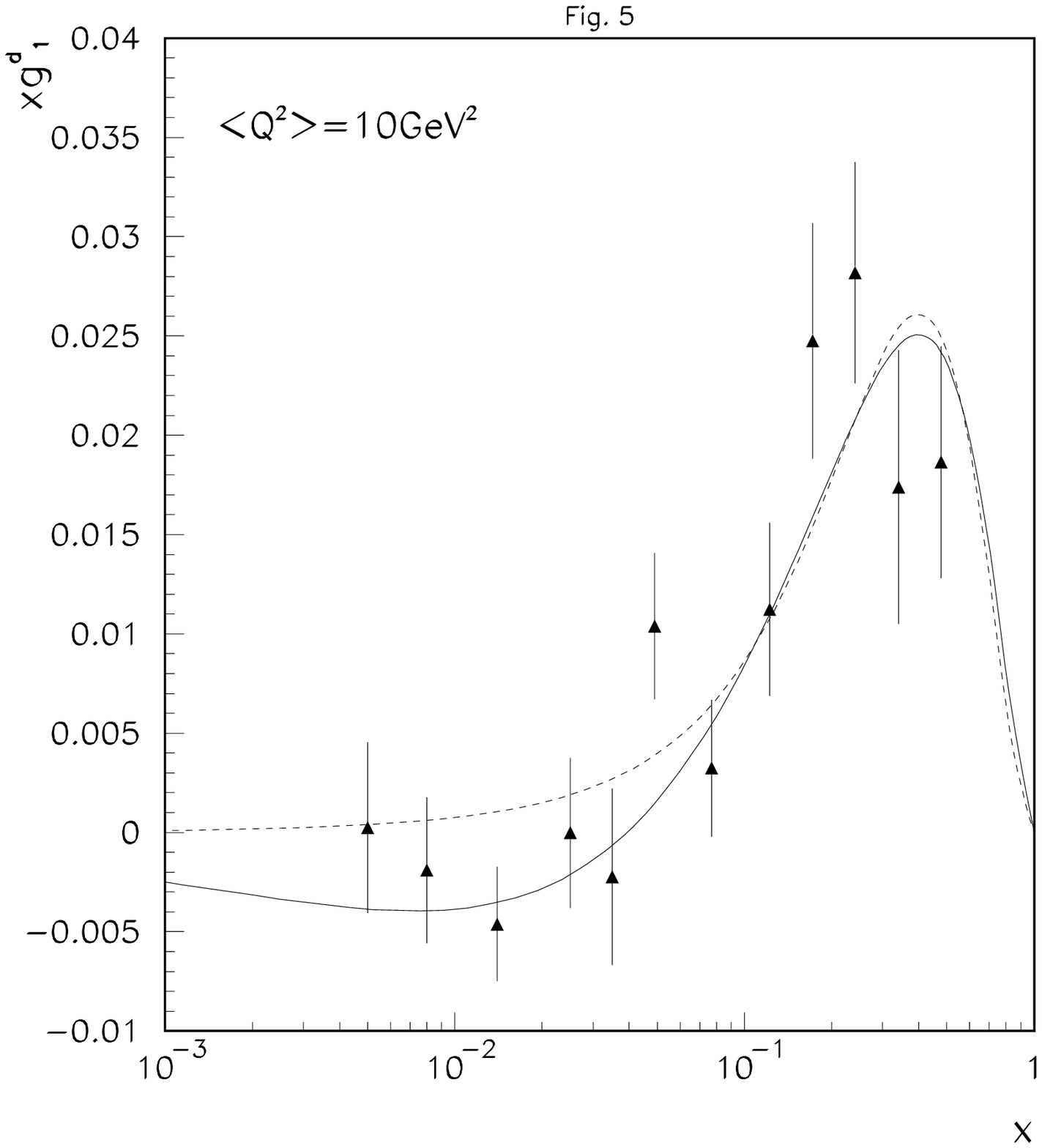,height=15cm}}
\label{fig:fig5}
\end{figure}

\end{document}